\documentstyle[aps,prd]{revtex}

\newcommand{\bq}{\begin{equation}}
\newcommand{\eq}{\end{equation}}
\newcommand{\bqn}{\begin{eqnarray}}
\newcommand{\eqn}{\end{eqnarray}}
\newcommand{\nb}{\nonumber}
\newcommand{\lb}{\label}

\begin{document} 
\title{  Collapsing Perfect Fluid in  Higher Dimensional Spherical 
Spacetimes} 
\author{Jaime  F. Villas da Rocha \thanks{E-mail: roch@on.br;
roch@dft.if.uerj.br}} 
\address{ Departmento de Astronomia Gal\'atica e
Extra-Gal\'atica,   Observat\'orio Nacional~--~CNPq,  Rua General Jos\'e
Cristino 77, S\~ao Crist\'ov\~ao, 
 20921-400 
Rio de Janeiro~--~RJ, Brazil }
\author{ Anzhong Wang\thanks{E-mail: wang@dft.if.uerj.br}}
\address{  Departamento de F\' {\i}sica Te\' orica,
Universidade do Estado do Rio de Janeiro,
 Rua S\~ ao Francisco Xavier 524, Maracan\~ a,
20550-013 Rio de Janeiro~--~RJ, Brazil}

\date{\today}

\maketitle

\begin{abstract}

The general metric for N-dimensional spherically symmetric and conformally
flat spacetimes is given, and all the homogeneous and isotropic 
solutions for a perfect fluid with the
equation of state  $p = \alpha \rho$  are found. These solutions are then used
to model the gravitational collapse of a compact ball. It is found that when
the collapse has continuous self-similarity, the formation of black holes
always starts with zero mass, and when the collapse has no such  a symmetry,
the formation of black holes always starts with a   mass gap.  
\end{abstract}

\vspace{.6cm}


\section{ Introduction}

Recently, we have studied the gravitational collapse of perfect fluid
in four-dimensional spacetimes \cite{WRS1997,RWS1999}, and found that when
solutions have continuous self-similarity, the formation of black holes
{\em always} starts with zero-mass, while when solutions have no  such a
symmetry it starts with a mass gap. The solutions with zero masses
actually represent naked singularities. Thus, if the Cosmic Censorship
Conjecture is correct \cite{Penrose1969}, it seems that Nature prohibits the
existence of solutions of the Einstein field equations with self-similarity. 

Quite recently,  there has been renewed interest in studying higher
dimensional spacetimes from the point of view of both Cosmology \cite{RS1998}
and gravitational collapse\cite{SH1996}. In particular, it was found that the
exponent $\gamma$, appearing in the mass scaling form of black holes,  depends
on the dimensions of the spacetimes \cite{GCD1999}.  

In this paper, we shall generalize our previous studies to the case of perfect
fluid in N-dimensional spherical spacetimes and will show that our previous
results for gravitational collapse, obtained in four-dimensional spacetimes 
\cite{WRS1997,RWS1999}, are also valid in N-dimensional spacetimes. The
solutions to be presented below can be also considered as representing
cosmological models. However, in this paper we shall not consider this
possibility and leave it to another occasion. The rest of the paper is
organized as follows: In Sec.$II$ we shall derive the general form of metric
for conformally flat spherical spacetimes with N-dimensions. As an
application, all the homogeneous and isotropic Friedmann-Robertson-Walker-like
solutions for a perfect fluid with the equation of state  $p = \alpha \rho$ 
are found for spacetimes with any dimensions. The solutions can be classified
into three different families, flat, open, and close,  depending on the
curvature of the $(N-1)$-dimensional spatial part of the metric, as that in
four-dimensional case. In Sec. $III$, two families of these solutions are
studied in the context of gravitational collapse, one   has continuous 
self-similarity, and the other   has neither continuous self-similarity nor
discrete self-similarity. In order to study gravitational collapse in
N-dimensional spacetimes, in this section  a definition for the mass function
is first given, whereby the location of the apparent horizons can be read off
directly. This definition gives the correct results for the static case
\cite{MP1986} and reduces to the four-dimensional one when $N = 4$
\cite{PI1990}. The paper is ended with Sec. $IV$, where our main conclusions
are derived. 

It should be noted that multidimensional spacetimes where all
dimensions are in the equal foot, like the ones to be considered here, are
not so realistic, as we are living in an effectively 4-dimensional spacetime,
so in principle one might expect that by dimensional reduction the
multidimensional  spacetimes should reduce to our 4-dimensional world. To have
such reduction possible, the dimensions should have different weights in any
realistic model. In this sense, the models considered in this paper are very
ideal.

\section{The General Homogeneous and Isotropic Solutions in N-dimensional
Spacetimes}

To  start with, let us consider the  general metric
for N-dimensional spacetimes with spherical symmetry \cite{Ei1925},
\bq
\lb{eq1}
ds^{2} = G(t, r) dt^{2} - K(t, r)\;\left(dr^{2}  
+ r^{2}d\Omega^{2}_{N-2}\right),
\eq
where  $\{x^{\mu}\}\equiv \{t, r, \theta^{2}, \theta^{3}, ...,
\theta^{N-1}\}\; (\mu = 0, 1, 2, ..., N-1)$ are  the usual
N-dimensional spherical coordinates,   $d\Omega^{2}_{N-2}$ is the line element
on the unit (N-2)-sphere, given by 
\bqn
\lb{eq2}
d\Omega^{2}_{N-2} &=& \left(d\theta^{2}\right)^{2} +
\sin^{2}(\theta^{2})\left(d\theta^{3}\right)^{2} + 
\sin^{2}(\theta^{2})\sin^{2}(\theta^{3})\left(d\theta^{4}\right)^{2}\nb\\
& & + ... + \sin^{2}(\theta^{2})\sin^{2}(\theta^{3}) ...
\sin^{2}(\theta^{N-2})\left(d\theta^{N-1}\right)^{2}\nb\\
& = &  \sum^{N-1}_{i = 2}{\left[
\prod^{i-1}_{j =2}\sin^{2}(\theta^{j})\right]
\left(d\theta^{i}\right)^2}.
\eqn
Then, it can be shown that the conformally flat condition 
$C_{\mu\nu\lambda\delta} =0$,  where $C_{\mu\nu\lambda\delta}$ denotes the
Weyl tensor, reduces to a single equation
\bq
\lb{eq14} 
D,_{rr} - \frac{D_{,r}}{r} = 0, 
\eq
where $D \equiv (G/K)^{1/2}$, and $(\;),_{r} \equiv \partial(\; 
)/\partial r$, etc. The   general solution of the above equation is given by
\bq 
\lb{eq15} 
D(t, r) = f_{1}(t) + f_{2}(t) r^{2}, 
\eq
where $f_{1}$ and $f_{2}$ are two arbitrary functions of $t$ only. Using the
freedom of coordinate transformations, it can be shown that there are
essentially only two different cases, one is $f_{1}(t) = 1, \;
f_{2}(t) = 0$, and the other is $f_{1}(t) \not= 0, \; f_{2}(t) = 1$. 
Thus, it is concluded that {\em the  general conformally flat
N-dimensional metric with spherically symmetry} takes  the form
 \bq
\lb{eq17}
ds^{2} = G(t, r) \left[ dt^{2} - h^{2}(t, r)\;\left(dr^{2} 
+ r^{2}d\Omega^{2}_{N-2}\right)\right],
\eq
where 
\bq
\lb{eq18}
h(t, r) = \left\{
\cases{1, \cr
{\left[f_{1}(t) + r^{2}\right]}^{-1},cr}
\right.
\eq
with $f_{1}(t) \not= 0$. In the following we shall refer solutions
with $h(t,r) = 1$ as Type $A$ solutions, and solutions with $h(t,r) =
[f_{1}(t) + r^{2}]^{-1}$ as Type $B$ solutions.  

When $f_{1}(t) = Const.$, say,
$f_{1}$, we can introduce a new radial coordinate $\bar{r}$ via the relation
\bq
\lb{eq19}
\bar{r} = \frac{r}{f_{1} + r^{2}},
\eq
then the metric (\ref{eq2}) becomes
\bq 
\lb{eq20} 
ds^{2} = G(t, r) \left( dt^{2} -  \frac{{d\bar{r}}^{2} }{1 -
4f_{1}\bar{r}^{2}} - \bar{r}^{2}d\Omega^{2}_{N-2}\right),\;\;\;
\left(f_{1}(t) = Const.\right).
\eq 
If we further set $G(t,r) = G(t)$, the above metric becomes the 
Friedmann-Robertson-Walker (FRW) metric but in N-dimensional spacetimes. 
Solving the corresponding Einstein field equations for a perfect
fluid with  the equation of state  $p = \alpha \rho$, we find
two classes of solutions, where $\rho$ denotes the
energy density of the fluid, $p$ the pressure, and  $\alpha$ is an arbitrary
constant. The details of the derivation of these solutions
were given in \cite{RW1999}, so in the following we should only present the
final form of the solutions.

{\bf Type $A$ solutions}.  In this case, the general solutions are
given by  
 \bq  
\lb{eq21} 
h(t, r) = 1,\;\;\;\;  G(t,r)  =  \left(1 - Pt  \right)^{2 \xi}, 
\eq  
where $P$ is a constant, which characterizes the strength of the spacetime
curvature. In particular, when $P = 0$, the spacetime becomes Minkowski. The
constant $\xi$ is a function of $\alpha$ and the spacetime dimension $N$, given
by   \bq 
\lb{eq22} 
\xi \equiv  
\frac{2}{(N-3)+ \alpha (N-1)}. 
\eq 
 The corresponding energy density and velocity of the fluid are given,
respectively, by 
\bqn
\label{eq23}
p &=& \alpha \rho  =  {3\kappa^{-1}(N-1)\alpha  \xi^2 P^2}{\left( 1 - Pt 
\right)^{-2(\xi+1)}  },\nb\\ u_0 & =& {\left(1 - Pt  \right)^{\xi}  },
\; \; \; \; \; \; u_{i}  = 0,\; (i =  1, 2, ..., N-1).
\eqn
This class of solutions belongs to the ones studied by Ivashchuk and Melnikov
in the context of higher dimensional Cosmology \cite{IM1989}. As a matter
of fact, setting $n = 0$ and $N_{0} = N-1$ in their general solutions, we
shall obtain the above ones. When $\alpha  = 0, (N-1)^{-1}$,
the above solutions reduce, respectively, to the one for a dust and radiation
fluid, which were also studied recently by Chatterjee and Bhui 
\cite{CB1990}. The $ \alpha = 0$ case was studied in the context of
gravitational collapse, too \cite{BSC1994}.    It can be shown that the 
curvature of the (N-1)-dimensional spatial part of the metric $t = Const.$ in
this case is zero, and the spacetimes correspond to the spatially flat FRW
model.

{\bf Type $B$ solutions}. In this case, the general  solutions are given by, 
\bq 
\lb{eq24}
h(t, r) = \frac{1}{f_{1} + r^{2}},\;\;\; G(t)  = 
\left[A \cosh(\omega t)
+ B \sinh(\omega t) \right]^{2\xi},
\eq
where $ \omega \equiv  2 \sqrt{-f_1}/ \xi$, $A$ and $B$ are integration
constants, and $\xi$ is  defined by Eq.(\ref{eq22}).
 The energy density and
velocity of the fluid now are given by
\bqn
\label{eq25}
p &=& \alpha  \rho  =  
12\kappa^{-1}N(N-1)\alpha f_{1}(A^{2} - B^{2})\left[A \cosh(\omega t) 
+ B \sinh(\omega t) \right]^{-2(1+\xi)}, \nb\\
u_0 & = & \left[A \cosh(\omega t) 
+ B \sinh(\omega t) \right]^{\xi},\;\;\;  u_{i}  = 0,\; (i =  1, 2, ... N-1).
\eqn
It can be shown that the  curvature of the (N-1)-dimensional spatial part of
the metric $t = Const.$  in the present case is different
from zero. In fact, when $f_{1} > 0$, the curvature is positive, and the
spacetime is spatially closed, and when $f_{1}< 0$, the curvature is negative,
and the spacetime is spatially open. The particular solution with $\alpha =
-(N-3)/(N-1)$ was found in \cite{CL1999}, given by Eqs.(49) and (50), but
its physical studies were excluded. As far as we know, the rest of the
solutions are new.    

It should be noted that the above solutions are valid for any constant
$\alpha$. However, in the rest of the paper we shall consider only the case
where $ -(N-2)^{-1} \le \alpha \le 1$, so that the three energy conditions,
weak, dominant, and strong,   are satisfied \cite{HE1973}. When $N
= 4$, these solutions reduce to the FRW solutions, which have been studied in
the context of gravitational collapse in \cite{WRS1997,RWS1999}. Therefore, in
the following we shall assume that $N \not= 4$.

\section {Gravitational Collapse of Perfect Fluid in N-dimensional Spacetimes}

To study the above solutions in the context of gravitational collapse, we
need first to define the local mass function. Recently, Chatterjee
and Bhui gave a definition, in generalizing the Cahill and Macvittie  mass
function  in four-dimensional spacetimes \cite{CM1973} to N-dimensional
spacetimes \cite{CB1993}. However, in
this paper we shall use the following definition for the mass function, 
\bq 
\label{eq27} 
1- \frac{2 m(t,r)} { B_{N}r_{ph}^{N-3}} = -g^{\mu
\nu}{r_{ph}}_{,\mu} {r_{ph}}_{,\nu},
\eq
where 
\bq
\lb{eq28}
B_{N} = \frac{\kappa \Gamma\left(\frac{N-1}{2}\right)}{2(N-2)\pi^{(N-1)/2}},
\eq
with $\Gamma$ denoting the gamma function, and $r_{ph}$ the geometric radius
of the (N-2)-unit sphere.  It can be shown that this definition in general is
different from that given by  Chatterjee and Bhui \cite{CB1993}, and  reduces
to the one usually used in four-dimensional  spacetimes when $N = 4$
\cite{PI1990}, and  yields the correct mass of black holes in N-dimensions for
the static spherical spacetimes \cite{MP1986}. 

In the study of gravitational collapse, another important conception is
the apparent horizon, the formation of which indicates the formation of
black holes. The apparent horizon in the present case is defined as the outmost
boundary of the trapped $(N-2)$-spheres  \cite{HE1973}. The location of
the trapped $(N-2)$-spheres is the place where the outward normal of the
surface, $ r_{ph} = Const.$, is null, i.e.,
\bq 
\lb{eq29}  
g^{\mu \nu}{r_{ph}}_{,\mu} {r_{ph}}_{,\nu} = 0.  
\eq 
Then, the mass
function on the apparent horizon is given by 
\bq  
\lb{eq30}   
M_{AH} =\left.
\frac{B_{N}}{2} r_{ph}^{N-3}\right|_{r = r_{AH}}, 
\eq 
where $r = r_{AH}$ is a solution of Eq.(\ref{eq29}), which corresponds to the
outmost trapped surface. In gravitational collapse $M_{AH}$ is usually taken as
the  mass of black holes \cite{Gu1997}.   With the above definition for the
mass function, let us study the main properties of the above two types of
solutions separately.

\subsection{ Type A solutions }

The mass function defined by Eq.(\ref{eq27}) in this case takes the form
\bq
\label{eq31}
m(t,r) = \frac{B_{N}}{2}\frac{\xi^{2}P^{2}r^{N-1}}{(1 - Pt)^{2 - \xi(N-3)}},
\eq
while Eq.(\ref{eq29}) has the solution  
\bq
\label{eq32}
r_{AH} = {1 \over \xi} {| 1- Pt | \over | P |},
\eq
which represents the location of the apparent horizon of the solutions. When 
$\xi = 1 $,  the apparent horizon
represents a null surface in the $(t,r)$-plane, and when $ 0 \le \xi <
1$, the apparent horizon  is spacelike, and when $\xi = 1$, it is null, while  
when $   \xi > 1$, it is  timelike. The spacetime is singular when
$t = 1/P$. This can be seen, for example, from the Kretschmann scalar, which
now is given by
\bq
\label{eq33}
{\cal{R}} \equiv
R^{\alpha\beta\gamma\delta} R_{\alpha\beta\gamma\delta}
  = 
 6\xi^{2} P^{4}\left\{N- 1+\xi^{2}
\left[ N- 2 + \sum_{A=1}^{N-3} (N-2-A) \right] \right\}
(1 - Pt)^{-4(1+\xi)}.
\eq
When $P > 0$, it can be shown that the singularity always hides behind the
apparent horizon, and when $P < 0$, the singularity is naked. In the latter
case, the solutions can be considered as representing cosmological models,
while in the former the solutions as representing the formation of black holes
due to the gravitational collapse of the perfect fluid. 
Substituting Eq.(\ref{eq32}) into Eq.(\ref{eq30}) we find that, as $t
\rightarrow + \infty$, the mass of the black hole becomes infinitely large. To
remend this shortage, one may follow \cite{WO1997,WRS1997,RWS1999} to cut the
spacetime along a timelike hypersurface, say, $r = r_{0}(t)$, and then join the
part with $r < r_{0}(t)$ with an asymptotically flat N-dimensional
spacetimes. From Eqs.(\ref{eq23})   we can see that the fluid is comoving with
the coordinates. Thus, the timelike hypersurface now
can be chosen as $r = r_{0} = Const.$ Then, it can be seen that at the
moment $t_{c} = - (P\xi r_{0}-1)/P$, the whole ball collapses inside
the apparent horizon, so the contribution of the collapsing ball to the total
mass of such a formed black hole is given by
\bq
\lb{mass1}
M_{BH}^{F} = m(t_{c}, r_{0}) = \frac{B_{N}}{2}
{\xi^{\xi(N-3)}r_{0}^{(N-3)(1+\xi)}}P^{\xi(N-3)}.
\eq
From the above expression we can see that the  mass is
proportional to $P$, the parameter that characterizes the strength
of the initial data of the collapsing ball. Thus, when the initial data is very weak
($P \rightarrow 0$), the mass of the formed black hole is very small
($M_{BH} \rightarrow 0$). In principle, by properly tuning the parameter
$P$ we can make it as small as wanted. 
 
It is interesting to note that this class of solutions admits a homothetic
Killing vector,
\bq 
\label{eq34} 
\zeta_{0}  =  - \frac{1- Pt}{ (1 +  \zeta) P},  \; \; \; 
\zeta_1    =   \frac{r}{ 1 +  \zeta}, \; \; \; 
\zeta_{i}    =   0, \; (i = 2,3, ..., N-1),
\eq 
which satisfies the conformal Killing equation, 
\bq
\lb{eq35}
\zeta_{\mu;\nu} + \zeta_{\nu;\mu} = 2 g_{\mu\nu}.
\eq
Introducing two new coordinates via the relations,
\bq
\lb{eq36}
\tilde{t} = \frac{(1 - Pt)^{\xi + 1}}{(1 + \xi)P},\;\;\;
\tilde{r} = r^{1 + \xi},
\eq
we find that the metric  can  be written  in an explicit self-similar form,
\bq  
\label{eq37} 
ds^2  =    {d\tilde t}^2 -   
\left[ { {(\xi +1)}^{-1/\xi} P x} \right]^{2\xi \over \xi +1} d{\tilde r}^2 
- \left[ {  {(\xi +1)} P} x \right]^{2\xi \over \xi +1}  
{\tilde r}^2 d\Omega^{2}_{N-2}, 
\eq
where $x \equiv \tilde{t}/\tilde{r}$ is the self-similar variable.

It is well-known that an irrotational 
``stiff" fluid ($\alpha = 1$) in
four-dimensional spacetimes is  energetically equal to a massless scalar
field \cite{TT1973}. It can be shown that this also the case for N-dimensional
spacetimes. In particular, for the above solutions with $\alpha = 1$,   the
corresponding massless scalar field $\phi$ is given by 
\bq   
\lb{eq38}  
 \phi  =\pm \left[\frac{N-1}{\kappa (N-2)}\right]^{1/2}\ln\left(1-Pt\right)
+\phi_{0},  
\eq 
where $\phi_{0}$ is a constant.

\subsection{ Type B solutions} 

The solutions in this case are given by  Eq.(\ref{eq24}). When
$f_{1} > 0$, the spacetime is close, and  to have the metric be real
the constant $B$ has to be imaginary.  The spacetimes are singular when,
\bq
\lb{eq39}
t|_{f_{1} > 0} = \frac{1}{|\omega|}{\mbox{arctan}}\left(\frac{|B|}{A}\right) +
2n\pi, \eq
where $n$ is an integer. When $f_{1} < 0$, the spacetime is singular only
when
\bq
\lb{eq40}
t|_{f_{1} < 0} = \frac{1}{\omega}{\mbox{arctanh}}\left(\frac{B}{A}\right).
\eq
Therefore, in the following we shall consider only the case where $f_{1} < 0$.
In this case, to have the energy density of the fluid be non-negative, we
need to impose the condition $B^{2} \ge A^{2}$. Then, the metric coefficient
$G(t)$ can be written as 
\bq
\lb{eq41}
G = \left(B^{2} - A^{2}\right)^{\xi}
 \sinh^{2\xi}[\omega(t_{0} - \epsilon t)],
\eq
 where  $\epsilon = sign(B)$, and $t_{0}$ is defined as 
$$
\sinh(\omega t_{0}) = \frac{A}{(B^{2} - A^{2})^{1/2}}.
$$
Clearly, the conformal factor $(B^{2} - A^{2})^{\xi}$ does
not play any significant role, without loss of generality, in the following we
shall set it to be one. If we further introduce a new radial coordinate via the
relation,
\bq
\lb{eq42}
\bar{r} = - \int{h(t,r) dr} = \frac{1}{a}\ln\left|\frac{a + r}{a - r}\right|,
\eq
where $ a \equiv (-f_{1})^{1/2}$, the corresponding metric takes the form,
\bq
\lb{eq43}
ds^2 = 
{\sinh}^{2 \xi} \left[ {2 \over \xi} \left( t_0 -\epsilon t \right) \right]
\left\{ dt^2 - d{ r}^2 - {{\sinh}^2(2  { r})\over 4 }
d\Omega^{2}_{N-2} \right\}. 
\eq
Note that in writing the above expression, we had set, without loss of
generality, $a = 1$, and dropped the bars from  $r$. From this metric, it can
be shown that it is not self-similar. In the following we shall use this form
of metric for the study of the Type B solutions. The corresponding mass
function and Kretschmann scalar are given, respectively, by 
\bqn  
\lb{eq44}
m(r,t) &=&   {B_{N} \over 2^{N-2} } \sinh^{N-1}(2  {r} )
\sinh^{\xi(N-3) -2} {\left[ {2\xi^{-1}} {( t_0 - \epsilon t)} 
\right]}, \nb\\
{\cal{R}}  &=&   \frac{96}{\xi^2} \left\{ (N-1) +\xi^2
\left[ N-2 + \sum_{A=1}^{N-3} (N-2-A) \right] \right\} 
\times 
\sinh^{-4 (\xi +1)}  
{\left[ {2 \xi^{-1}} {\left( t_0 - \epsilon t\right)} 
\right]},
\eqn
while the apparent horizon is located at
\bq
\lb{eq45}
r = r_{AH} \equiv { \xi^{-1} } (t_0 - \epsilon t).
\eq
From the above equations, we can see that  the solutions
are singular  on the hypersurface $t =  \epsilon t_{0}$. When $
\epsilon = +1$, the   singularity is hidden behind the apparent horizon, and
the solutions  represent  the formation of black holes from the gravitational
collapse of the fluid. When $ \epsilon =  - 1$,  the singularity is naked, the
solutions can be considered as representing cosmological models or   white
holes.  As in the type A case, the mass   of such formed black holes also
diverges at the    limit $t \rightarrow + \infty$. Thus, to have finite masses
of black holes, we may also make a ``surgery"  to the spacetimes. Since the
fluid is comoving with the coordinates, too. Without loss of generality, we
may also choose the boudary as $r = r_{0}$. Then, it can be shown that
the contribution of the collapsing ball to the total mass of black hole
now is given by
\bq
\lb{mass2}
M^{F}_{BH} \equiv m_{AH}(\tau_{AH}, r_{0}) = 
\frac{B_{N}}{2^{N-2}} \sinh^{(N-3)(\xi +1)}(2r_{0}).
\eq
From the above expression we can see that for any given non-zero $r_{0},\;
M^{F}_{BH}$ is always finite and non-zero. Thus, in the present case black
holes start to form with a mass gap. 

Finally we note that, similar to the last case, the solution with $\alpha =1$
also corresponds to a massless scalar field with the scalar field being given
by  
\bq
\lb{eq46}
\phi = \pm
\left[\frac{N-1}{\kappa(N-2)}\right]^{1/2}\ln\left\{\tanh[(N-2)(t_{0} -
\epsilon t)]\right\} + \phi_{0},\; (\alpha = 1). 
\eq

\section{ Concluding Remarks}

The general form of metric for N-dimensional spherically symmetric
and conformally flat spacetimes was found. As an application of it, all
the Friedmann-Robertson-Walker-like  solutions for a perfect fluid with an
equation of state  $p = \alpha \rho$  were given. These solutions were then
used to model the gravitational collapse of a compact ball. It was shown that
when the collapse has continuous self-similarity, the formation of black holes
always starts with zero mass, and when the collapse has no such a symmetry,
the formation of black holes always starts with a finite non-zero mass. This
is the same as that obtained in the study of the problem in four-dimensional
spacetimes \cite{WRS1997,RWS1999}. Thus, they   provide further evidences to
support the speculation that {\em the formation of black holes always starts 
with zero-mass for the collapse with self-similarities}.

\acknowledgments

We would like to express our gratitude to S.K. Chatterjee for
valuable discussions. The financial assistance from CAPES (JFVR), CNPq (AW)
and FAPERJ (AW) is gratefully acknowledged.


\begin{thebibliography}{99}


\bibitem{WRS1997} A.Z. Wang, J.F.Villas da Rocha, and N.O. Santos, 
 Phys. Rev. {\bf D56}, 7692 (1997).

\bibitem{RWS1999} J.F.Villas da Rocha, A.Z. Wang, and N.O. Santos, 
 Phys. Lett. {\bf A255}, 213 (1999).

\bibitem{Penrose1969} R. Penrose, Rivista Del Nuovo Cimento, (Numero Special)
{\bf 1}, 252 (1969).

\bibitem{RS1998} S.~Ferrara, M.~Porrati, and A.~Zaffaroni,  Lett. Math. Phys.
{\bf   47}, 255 (1999);   A.~Khavaev, K.~Pilch, and N.~P. Warner, ``{\em New
vacua of gauged {N=8}   supergravity in five-dimensions},'' {{\tt
hep-th/9812035}}; L.~Randall and R.~Sundrum, ``{\em A large mass hierarchy from
a small extra  dimension},'' {{\tt hep-ph/9905221}};   ``{\em An alternative to
compactification},''   {{\tt hep-th/9906064}};
N.~Arkani-Hamed, S.~Dimopoulos, G.~Dvali, and N.~Kaloper, ``{\em Infinitely
large   new dimensions},'' {{\tt hep-th/9907209}};
A.~E. Nelson, ``{\em A new angle on intersecting branes in infinite extra
dimensions},'' {{\tt hep-th/9909001}}.    

\bibitem{SH1996} J. Soda and K. Hirata, Phys. Lett. {\bf B387}, 271 (1996);
A.V. Frolov, Class. Quantum Grav. {\bf 16}, 407 (1999).

\bibitem{GCD1999} D. Garfinkle, C. Cutler, and G.C. Duncan, Phys. Rev. {\bf
D60}, 104007 (1999).

\bibitem{MP1986} R.C. Myers and M.J. Perry, Ann. Phys. (N.Y.), {\bf 172}, 304
(1986).

\bibitem{PI1990}E. Poisson and W. Israel,  Phys. Rev. {\bf D41}, 1796 (1990).

\bibitem{Ei1925} L.P. Eisenhardt, {\em Riemannian Goemetry} (Princeton
University Press, Princeton, 1925).

\bibitem{RW1999} J.F. Villas da Rocha and A.Z. Wang,  ``{\em Gravitational
Collapse of Perfect Fluid in N-Dimensional Spherically Symmetric Spacetimes},"
{\tt gr-qc/9910109}, preprint (1999).

\bibitem{IM1989} V.D. Ivashchuk and V.N. Melnikov, Phys. Lett. {\bf A135}, 465
(1989); Gravitation $\&$ Cosmology, {\bf 1}, 133 (1995).

\bibitem{CB1990} S. Chatterjee and B. Bhui, Mon. Not. R. astr. Soc. {\bf
247}, 57 (1990).

\bibitem{BSC1994} A. Banerjee, A. Sil, and S. Chatterjee, Astrophys. J. {\bf
422}, 681 (1994).

\bibitem{CL1999} S. Capozziello and G. Lambiase,   Gravitation $\&$
Cosmology, {\bf 5}, 131 (1999).


\bibitem{HE1973} S.W. Hawking and G.F.R. Ellis, {\em The Large Scale
Structure of Space-Time} (Cambridge University Press, Cambridge,
1973) pp. 88-96.

\bibitem{CM1973} M. E. Cahill and G. C. McVittie J. Math. Phys., {\bf 11}
1382 (1970).

\bibitem{CB1993} S. Chatterjee and B. Bhui, Inter. J. Theor. Phys. {\bf
32}, 671 (1993).


\bibitem{Gu1997} C. Gundlach, Adv. Theor. Math. Phys. {\bf
2}, 1 (1998), $gr-qc/9712084$ (1997).
 
\bibitem{WO1997} A.Z. Wang and H.P. de Oliveira, Phys. Rev. 
{\bf D56}, 753 (1997).

\bibitem{TT1973} R. Tabensky and A.H. Taub, Common. Math. Phys. {\bf 29}, 61
(1973).


\end{thebibliography}
\end{document}